\documentclass[pdflatex,sn-mathphys-num]{sn-jnl}

\usepackage{gensymb}
\usepackage{graphicx}%
\usepackage{multirow}%
\usepackage{amsmath,amssymb,amsfonts}%
\usepackage{amsthm}%
\usepackage{mathrsfs}%
\usepackage[title]{appendix}%
\usepackage{xcolor}%
\usepackage{textcomp}%
\usepackage{manyfoot}%
\usepackage{booktabs}%
\usepackage{algorithm}%
\usepackage{algorithmicx}%
\usepackage{algpseudocode}%
\usepackage{listings}%

\begin{document}

\title[Recent Results on the Tetraneutron]{Recent Results on the Tetraneutron}


\author*[1]{\fnm{Thomas} \sur{Faestermann}}\email{thomas.faestermann@ph.tum.de}
\equalcont{These authors contributed equally to this work.}

\author[1]{\fnm{Roman} \sur{Gernh\"auser}}\email{roman.gernhaeuser@ph.tum.de}
\equalcont{These authors contributed equally to this work.}


\affil*[1]{\orgdiv{TUM School of Natural Sciences, Department of Physics}, \orgname{Technical University of Munich}, \orgaddress{\street{James-Franck-Str. 1}, \city{Garching}, \postcode{85748}, \country{Germany}}}




\abstract{We describe the recent experiments which claimed an observation of a tetraneutron signal. Production reactions like transfer, knockout, fragmentation or photodisintegration have been used at very different experiments and facilities to form systems just made of neutrons. As a possible explanation of the partly contradicting results we suggest that some observed the bound ground state and some 
an unbound but still correlated  state of the four neutrons at different exitation energy.
We also refer to some of the theoretical works.}




\maketitle

\section{Introduction}\label{sec1}

The experimental observation of a neutron cluster, a trineutron ($^3n$) or a tetraneutron ($^4n$), would not only show a really exotic nuclear system, but its properties, energy and width, would also impose constraints on the nuclear force in theoretical calculations which so far were not evident from state of the art experiments. If bound tetraneutrons existed, they certainly would be present in neutron stars and would influence their properties
but also could have exciting applications as the reactions of uncharged nuclei are not hindered by Coulomb repulsion. Since about seventy years experimental searches for tetraneutrons have been undertaken. But only in this century, indications and evidences for the production (and identification) of - bound and unbound - tetraneutron systems have been reported. These recent experiments will be briefly described, summarized and the results discussed in this little review. For experiments until 2016 and a more detailed review on theoretical work we refer to the review by Marqu\'es and Carbonell \cite{2021EPJA...57..105M}.

\section{Recent experiments}\label{sec2}




In the following, we describe the recent experiments in chronological order of publication.

\subsection{Fragmentation of $^{14}$Be at GANIL, France}
The first report in this century on a possible tetraneutron identification was by Marqu\'es et al.
\cite{PhysRevC.65.044006}. 
At GANIL they have fragmented a $^{14}$Be secondary beam of 35 MeV/nucleon on a C target. The charged breakup particles were detected in the forward direction with a Si-CsI telescope, which also yielded a particle identification signal in addition to the energy information. For the detection of neutral particles they used a 90 module liquid scintillator array about 5 m from the target. Neutrons were distinguished from $\gamma$ rays by pulse-shape analysis. 
The neutron energy $E_n$ was measured via their time-of-flight assuming a single neutron hitting the detector. The pulse-height signal in the scintillator yielded the energy of the recoil proton $E_p$ from the organic detector material that was compared with the neutron energy $E_n$. They found some seven events with an unusually large ratio $E_p/E_n \ge 1.4$, indicating an energy deposition larger than from a single neutron's recoil. All of these were coincident with identified $^{10}$Be events. They assigned these events to tetraneutrons which had survived the flight time to the neutron detectors of about 100 ns and therefore should be particle-stable. 
Later-on, the authors had to admit \cite{marques2005possible} that abnormally high signals in the neutron detector could also occur if more than one neutron from a tetraneutron decaying in flight were hitting a single detector module. 
For a decay into such a narrow cone they considered a maximum decay energy for $^4 n\Rightarrow 4n$ of 
$E\le 2.0$ MeV above the threshold. Unfortunately, later experiments at GANIL could not collect sufficient statistics to prove the observation \cite{2024FBS....65...37M}.

\subsection{Double charge exchange with $^{8}$He on $^{4}$He at RIKEN, Japan}
 The Radioactive Ion Beam Factory (RIBF) at RIKEN, Japan, is presently the most powerful facility for experiments using light secondary ions. The first use of a radioactive $^8$He $(T_{1/2}=$119 ms) beam in a search for a tetraneutron state at RIBF was by Kisamori et al. \cite{2016PhRvL.116e2501K} investigating the $^4$He($^8$He,$^8$Be)$^4n$ reaction at a beam energy of 186 MeV/nucleon. 
This can be viewed as a double-charge exchange reaction or as an $\alpha$-transfer reaction. The momenta of the two $\alpha$-particles from the decay of the $^8$Be ejectiles were measured with the magnetic spectrometer SHARAQ at $0\degree$. 
From these the missing mass of the rest (tetraneutron) can be reconstructed. They found a resonance with four events on negligible background, corresponding to a significance of $4.9\sigma$. 
The peak is at an energy of $0.83\pm 0.65(stat)\pm 1.25(syst)$ MeV above the four-neutron threshold. For the width they only could give an upper limit of $\Gamma \le 2.6$ MeV (FWHM).

\subsection{3-proton pickup from $^{7}$Li at the Technical University of Munich, Germany}
In Munich we attacked the tetraneutron problem with a stable beam. We chose the reaction $^7$Li$(^7$Li,$^{10}$C)$^4n$ with a 46 MeV $^7$Li beam \cite{2022PhLB..82436799F}. The center-of-mass energy is just 4.8 MeV above the $4n$ threshold.
Because this is a binary reaction it is sufficient to identify the $^{10}$C ejectile and to measure its momentum in order to deduce the energy of the tetraneutron. We used the high-resolution Q3D magnetic spectrograph at $7.75\pm 1.75 \degree$ equipped in the focal plane with a position sensitive single wire proportional counter plus an array of position sensitive Si detectors \cite{2018NPNDollinger}. From the energy loss in the wire detector and the residual energy in the Si detectors we could identify the ejectiles. As targets we used enriched $^7$Li$_2$O on a thin carbon backing.
In the energy spectrum of identified $^{10}$C ejectiles we detected two peaks. One at a $^{10}$C kinetic energy of 23.0 MeV corresponds to the $^{16}$O($^7$Li,$^{10}$C)$^{13}$B reaction, expected due to the composition of the target material. However, the other peak at 20.8 MeV, which has a significance of more than $3\sigma$, can only originate from a reaction on the $^7$Li. This energy corresponds to an excitation energy in the $^{10}$C$+4n$ system of 2.93 MeV. But the observed width corresponds to a width of a resonance of less than 0.24 MeV (FWHM). For a tetraneutron with an energy of 2.93 MeV available for the decay into four neutrons this width seems far too small. Therefore, we attributed this peak to an excited ejectile $^{10}$C$^*$ (E$^*(2^+)=3.354$ MeV) and a bound tetraneutron with a binding energy E$_B=0.42\pm 0.16$ MeV. In a second experiment we positioned the Q3D at $5.0\pm1.5\degree$ to prove the kinematically different energy shift of the two peaks due to the different target masses A=16 and A=7.
The result was now a broad peak composed of the two previous peaks, exactly as expected. 

\subsection{$\alpha$-knockout from $^{8}$He at RIKEN, Japan}
The group around M. Duer \cite{2022Natur.606..678D} used a 156 MeV/nucleon $^8$He beam to investigate the reaction $^8$He$(p,p\alpha)$ in inverse kinematics. The aim of the experiment was to scatter the proton and the $\alpha$-particle contained in the $^8$He quasi-elastically at a large momentum transfer and to impart only a negligible recoil momentum to the remaining four neutrons.
For a precise vertex reconstruction and particle identification of the charged reaction products $p$ and $\alpha$ an array of Si-strip detectors was used. They determined their momenta with the SAMURAI magnetic spectrometer and a drift chamber for tracking in its focal plane. In addition, two scintillator walls measured their deposited energy and time-of-flight. With these informations they reconstructed on an event-by-event basis the missing mass and the energy of the $4n$ system. In the missing mass spectrum for the four neutrons they observe, in addition to a non-resonant continuum, a peak of high significance which is well represented by a Breit-Wigner resonance. Its energy and width are $E_{4n}=2.37\pm0.38(stat.)\pm0.44(syst.)$ MeV and $\Gamma=1.75\pm0.22(stat.)\pm0.30(syst.)$ MeV.

\subsection{Photodisintegration of $^{209}$Bi in Yerevan, Armenia}
A team at the Alikanyan National Science Laboratory in Yerevan, Armenia \cite{2023JConP..58....6K} used Bremsstrahlung from electron beams to disintegrate $^{209}$Bi. After irradiation of a pure $^{209}$Bi target with photons they measured the $\beta$-delayed $\gamma$-radiation and the corresponding decay curves of the two most abundant transitions populated in the $\beta^+$/EC-decay of $^{205}$Bi (T$_{1/2}=14.91$ d \cite{KONDEV20201}). At a nominal electron energy of 30 MeV - that is 0.50 MeV above the threshold for the $^{209}$Bi($\gamma,4 n)^{205}$Bi reaction - they observed a cross section, weighted with the Bremsstrahlung spectrum, of $\sigma_w=0.442(0.048) \mu b$. 
At 28 MeV, no signal of the $^{205}$Bi production was obseved, with an upper limit of the weighted cross section of $\sigma \le 0.2 \mu b$. They compared the 30 MeV cross section for the emission of four neutrons with calculations using the TALYS1.9 and FLUKA models taking into account the energy spread of the electron beam of $2.5 \%$, with the result that the calculations underestimate the observed cross section at 30 MeV beam energy by more than an order of magnitude. They concluded that a correlated emission of four neutrons might explain this discrepancy because it is not included in the models. In their publication  \cite{2023JConP..58....6K} they propose, that the observed resonance structure from \cite{2022Natur.606..678D} could be a good explanation, if this is not caused by the reaction mechanism as postulated by Lazauskas et al. \cite{2023PhRvL.130j2501L}. However, if we plot into their calculated photon spectrum the resonance curve measured by Duer et al. \cite{2022Natur.606..678D} (see Fig.~\ref{fig:KD01}) we note only a minuscule overlap, even with the photon spectrum from 30 MeV electrons. A much better overlap of the spectrum from 30 MeV electrons can be observed with the bound tetraneutron claimed by Faestermann et al. \cite{2022PhLB..82436799F}.

\subsection{Proton and $\alpha$-transfer from $^{8}$He at JINR, Dubna, Russia}
Researchers at the ACCULINNA-2 facility in Dubna, Russia have also used a radioactive $^8$He beam with an energy of 26 MeV/nucleon to investigate the tetraneutron
\cite{PhysRevC.111.014612}. They used two reactions: $\alpha$-transfer from $^8$He in the $^2$H($^8$He,$^6$Li)$^4n$ reaction and proton transfer $^2$H($^8$He,$^3$He)$^7$H with subsequent decay $^7$H$\rightarrow ^3$H+$^4n$. For the detection and identification of the charged particles they used $\Delta$E-E-E telescopes consisting of Si strip detectors. Neutrons were detected in a large array of stilbene scintillators at a distance of 2 m from the target in forward direction. From the charged particle energy and angle they reconstructed the missing mass and the energy $E_{4n}$ in the $4n$ frame. Neutrons were distinguished from $\gamma$-rays using pulse-shape discrimination and the neutron energy was determined by their time-of-flight. In both reactions they found in the events coincident with a neutron detection a hump in the $E_{4n}$ spectrum which cannot be explained by the $4n$ phase space. The energies of these humps for the $4n$-system are $3.5\pm 0.7$ MeV and $3.2\pm 0.8$ MeV, respectively.

\section{Theoretical approaches to the tetraneutron}
The claim of a bound tetraneutron by the experiment of Marqu\'es et al.\cite{PhysRevC.65.044006} triggered a number of theoretical papers. The most definitive statement came from Steven Pieper who did Greens function Monte-Carlo calculations \cite{2003PhRvL..90y2501P}: "This means that, should a recent experimental claim of a bound tetraneutron be confirmed, our understanding of nuclear forces will have to be significantly changed."
As a consequence of the experiment of Kisamori et al.\cite{2016PhRvL.116e2501K} there was a large activity of theoreticians. With calculations using the no-core shell model several groups \cite{Shirokov.PhysRevLett.117.182502,2017PhRvL.119c2501F,2019PhRvC.100e4313L,2019PPN....50..537M} obtained tetraneutron resonances with excitation energies between about 1 MeV and 7 MeV and widths in excess of 1 MeV. Gandolfi et al. \cite{2017PhRvL.118w2501G} used chiral effective field theory and found a tetraneutron resonance at 2.1 MeV. However, continuum calculations \cite{2019PhRvC.100d4002D,Higgins.PhysRevLett.125.052501} did not result in a resonance for the tetraneutron system. In addition, there are a large number of theoretical papers, e.g. \cite{2016PhRvC..93d4004H,2017PTEP.2017g3D03L,2019PhRvC.100d4002D}, we certainly cannot mention all, which definitely deny the possibility of a resonance in the four-neutron system without varying the parameters of the nuclear potential too much to still be consistent with other observables. As a response to the Duer et al. experiment \cite{2022Natur.606..678D}, Lazauskas et al. \cite{2023PhRvL.130j2501L} interpreted the low-energy bump in the alpha knockout from $^8$He \cite{2022Natur.606..678D} as originating from the final-state interaction of two dineutrons in the exit channel. 

However, there are discrepancies between experiment and theory also in other light nuclear systems. Just recently a measurement of a low-lying resonance in $^6H$ was reported \cite{PhysRevLett.134.162501} using the $^7Li(e,e'p\pi ^+)^6H$ reaction. They find a resonance at $E_r =2.3 (0.5)(0.4)$ MeV above the threshold for disintegration into $^3H+3n$ and a width $\Gamma = 1.9 (1.0)(0.4)$ MeV, consistent with all other experimental results. However, the calculated values of Hiyama, Lazauskas and Carbonell \cite{2022PhLB..83337367H} are $E_r\approx
 10$ MeV and $\Gamma\approx 4$ MeV.

The most recent experimental papers \cite{2022PhLB..82436799F,2022Natur.606..678D} also triggered completely different theoretical approaches that were not on the shortlist of the problem so far. Energy density functionals (EDF) with density and momentum-dependent 2-body and 3-body interaction functionals are known to be a very successful concept for describing nuclei over a wide mass range. 
The Giessen EDF \cite{2019EPJA...55..238L,2018EPJA...54..170A} 
is derived microscopically from a G-matrix based on the Bonn nucleon-nucleon potential and supplemented by 3-body interactions of Urbana-type, the latter being well tested in 
Green's function Monte Carlo
calculations for light nuclei.  The Giessen EDF has been used to describe infinite matter and heavy nuclei (e.g. \cite{2019EPJA...55..238L,2018EPJA...54..170A}) but never was used for very light nuclei. 
For few-body calculations H. Lenske (Univ. Giessen, in Ref. \cite{2024Lenske}) started by enclosing N nucleons in a large spherical cavity to remove center-of-mass effects. 
Hartree-Fock-Bogoliubov theory was used to generate iteratively a self consistent mean-field. Only the medium range attraction was slightly tuned on a one percent level to reproduce binding energies and radial distributions of the well known isotopes $^{4,6,8}$He. Without any
other adjustment this approach generates for a four-neutron system a mean field with a strongly repulsive
core, reminiscent of Pauli-repulsion, and an attractive minimum at around 1.4 fm, roughly -30 MeV deep. Enlarging systematically  the cavity radius, at about 120 fm the $4n$ systems approaches the binding threshold. Finally, for $R \ge 140$ fm the system condenses into a bound state with a binding energy of +140 keV, remaining stable under further increases of the radius. Inspection of the wave function reveals  a compact s-wave core, essentially given by a correlated two neutron state. The core is immersed into a widely distributed component of  contributions up to d-waves, leading to a total rms-radius of the $4n$ system close to 45 fm. Thus, the typical dimensions of stable and known exotic nuclei are by far exceeded.
The density distribution looks like an extreme halo case. The halo part apparently plays a key role in stabilizing the $2n$ core. Even so these results are still preliminary this extreme structure would explain why few-body and  shell model like theories confined to several oscillator shells can hardly describe the $4n$ system, but also why this dilute
state could not be seen in Ref. \cite{2022Natur.606..678D,PhysRevC.111.014612}. 

\begin{figure}[t]
\begin{center}
	\includegraphics[width=8.5cm]{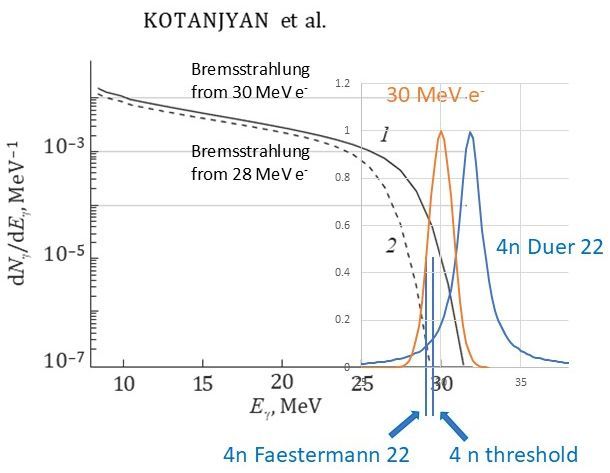}
	\caption{Copy of the calculated photon spectra for 30 MeV and
28 MeV electron beams (adopted from \cite{2023JConP..58....6K} with permission of T. V. Kotanjyan) . The electron spectrum for 30 MeV is shown in orange.The resonance curve of Duer et
al. \cite{2022Natur.606..678D} is drawn in blue. The energies necessary for independent four-neutron emission and for emission of a bound tetraneutron
as claimed by Faestermann et al. \cite{2022PhLB..82436799F} are indicated by vertical bars.}
	\label{fig:KD01}
\end{center}
\end{figure}

\begin{figure}[t]
\begin{center}
	\includegraphics[width=8.5cm]{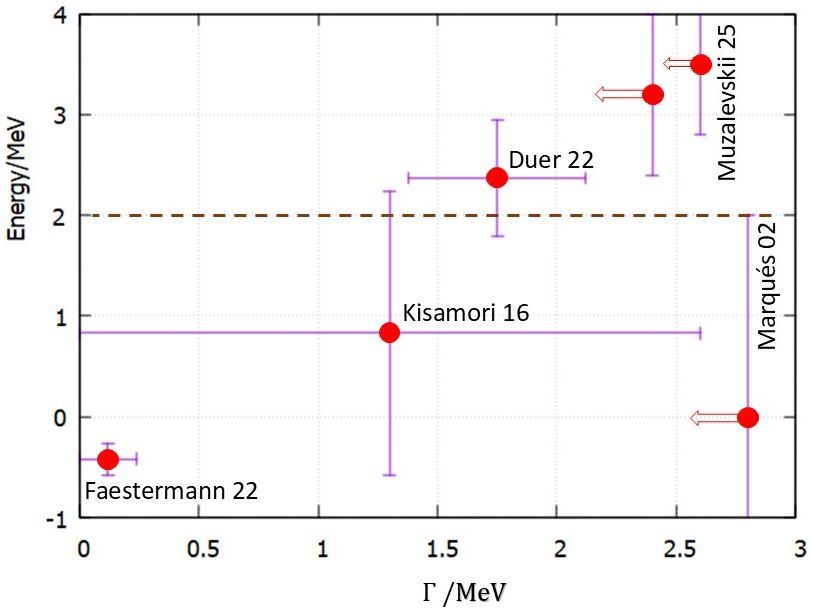}
	\caption{Results on a tetraneutron signal with respect to the width $\Gamma$ and energy of the $4n$ system. The results on the right, Marqu\'es02 \cite{PhysRevC.65.044006,marques2005possible} and Muzalevskii25 \cite{PhysRevC.111.014612} have no information on $\Gamma$. Kisamori16 \cite{2016PhRvL.116e2501K} and Faestermann22 \cite{2022PhLB..82436799F} have upper limits for the width, whereas Duer22 \cite{2022Natur.606..678D} has also a lower limit for the width.}
	\label{fig:Res01}
\end{center}
\end{figure}

\section{Synthesis of all experimental results}
We certainly would agree that a physical phenomenon is only understood, when there is a theory describing it. On the other hand, if there appears a conflict between theory and experiment, experimentalists have to make sure that the observed effect is indeed reproducible and certain, without being influenced by theoretical expectations. Additional experiments have been done already at RIBF and wait for the full analysis, like the proton knockout from $^8$He with the subsequent breakup of $^7$H into $^3$H and $^4n$  \cite{2021FBS....62..102H} and direct detection of the $^4n$ decay. Or a repeated double charge exchange experiment $^4$He($^8$He,2$\alpha)^4n$ \cite{2023NPNew..33c..15S} and a repetition of the $\alpha$-knockout \cite{2022Natur.606..678D}, also with a much larger multi-neutron detection efficiency. For the photodisintegration of $^{209}$Bi \cite{2023JConP..58....6K} more experiments at intermediate electron energies are being done.

Here we want to try to understand the presently available experimental information.
In Fig.~\ref{fig:Res01} we have summarized the results of the experiments described above with respect to the observed width $\Gamma$ and energy available for the decay into four neutrons. Only the experiment of Kotanjyan et al. \cite{2023JConP..58....6K} cannot specify the energy, but as it may be concluded from Fig.~\ref{fig:KD01} a state very close below or above  threshold seems to explain best the cross section difference between electron energies of 28 and 30 MeV. The result of Marqu\'es et al. \cite{PhysRevC.65.044006,marques2005possible} is only an upper limit for the energy and, as the results of Muzalevskii et al.  \cite{PhysRevC.111.014612}, gives no information on the width. Kisamori et al. \cite{2016PhRvL.116e2501K} and Faestermann et al. \cite{2022PhLB..82436799F} have upper limits for the width, whereas Duer et al. \cite{2022Natur.606..678D} has also a lower limit for the width. Concerning the energies Duer et al. \cite{2022Natur.606..678D} and Muzalevskii et al. \cite{PhysRevC.111.014612} are compatible with a weighted mean of $2.92\pm0.39$ MeV. All three results are certainly not compatible with the very precise, negative, value of Faestermann et al. \cite{2022PhLB..82436799F}. The values of Marqu\'es et al. \cite{PhysRevC.65.044006,marques2005possible} and Kisamori et al. \cite{2016PhRvL.116e2501K} may be compatible with both. Thus, Faestermann et al., on the one hand, and Duer et al. and Muzalevskii et al., on the other, cannot have observed the same state of the tetraneutron. However, could it not be that Faestermann et al. have observed the ground state and Duer et al. and Muzalevskii et al. the first excited state? We would expect the ground state to be formed from two neutrons in the $s_{1/2}$ shell and two neutrons in the $p_{3/2}$ shell coupled to spin zero. The first excited state would then have the two $p_{3/2}$ neutrons coupled to $2^+$. The energy difference would be $E^*=3.3\pm0.4$ MeV. 
For the four even-even neighbours with $N=4$: $^6$He, $^8$Be, $^{10}$C and $^{12}$O the average excitation energy of the $2^+$-state is $E^*=2.5\pm0.7$ MeV(RMS). (The uncertainty is the root-mean-square deviation.) In the target nucleus $^7$Li of the experiment of Faestermann et al.\cite{2022PhLB..82436799F} the neutrons are already in the lowest $0^+$ configuration, its spin is only determined by the protons which are removed in the reaction. In addition, low energy transfer reactions do not rely on a sudden process where the structural overlap of initial and final states are especially important but go through a relatively long compound phase mixing single particle wave functions.
However, in the $\alpha$-knockout as well as in the $\alpha$-transfer reactions it can be assumed that the $\alpha$ is formed from two neutrons in the $s_{1/2}$ shell which are removed.  Therefore, only $p_{3/2}$ neutrons would remain and it seems quite likely that the four neutrons in the final channel are in a $(p_{3/2})^2\ 2^+$ state. We have argued in a similar way before \cite{2022arXiv220710542F}. Or, to argue with radial dimensions, the $^8$He matter radius is with 2.49(4) fm \cite{2005EPJAS..25..215K} considerably larger than the charge radius of 1.959(16) fm \cite{PhysRevLett.99.252501,PhysRevLett.108.052504}  (Note that the charge radius of $^4$He is only 1.681(4) fm \cite{PhysRevC.77.041302}, but in $^8$He it can move around the common center of gravity of $^4$He and the four extra neutrons.). That means that the additional four neutrons form a halo around the strongly bound $\alpha$-particle. When the latter is removed there may be only a small overlap between the halo and the ground state configuration of the tetraneutron.
Duer at al.\cite{2022Natur.606..678D} implemented the $^8$He wave function in the initial state  and derived a quite large overlap with the $\alpha+4n$ configuration needed for the knockout. Nevertheless, this does not allow a conclusion on the preferred $4n$ configuration that will be populated. 

\section{Conclusions}

We have tried to find a common explanation for the recent experimental observations of a tetraneutron signal by assuming that some see the bound ground state and others the unbound resonance of the first excited state. This view, however, is in conflict with nearly all theoretical results published. Therefore, it is of utmost importance that more experimental efforts try to verify or falsify the described experimental results. If these indications of a bound state of four neutrons get manifest, we will have to search for ways to accommodate this fact in the models of the strong interaction.
\bigskip

\bmhead{Acknowledgements}
We are grateful to Horst Lenske and Tigran Kotanjyan for a critical reading of the manuscript.



\section*{Declarations}


\begin{itemize}
\item Funding

This work was supported by the Excellence Cluster ORIGINS from the German
Research Foundation DFG (Excellence Strategy
EXC-2094–390783311).

\item Author contribution

Both authors have contributed equally.
\end{itemize}



\bibliography{literature}

\end{document}